\documentclass{elsart}   

\usepackage{graphicx}   
   
\begin{document}   
\begin{frontmatter}   
\title {Correlation of Beam Electron and LED Signal Losses under Irradiation  
and Long-term Recovery of Lead Tungstate Crystals}   
\author[IHEP]{V.A.~Batarin},   
\author[FNAL]{J.~Butler},      
\author[IHEP]{A.M.~Davidenko},   
\author[IHEP]{A.A.~Derevschikov},   
\author[IHEP]{Y.M.~Goncharenko},   
\author[IHEP]{V.N.~Grishin},   
\author[IHEP]{V.A.~Kachanov},   
\author[IHEP]{A.S.~Konstantinov},   
\author[IHEP]{V.I.~Kravtsov},      
\author[UMN]{Y.~Kubota},   
\author[IHEP]{V.S.~Lukanin},   
\author[IHEP]{Y.A.~Matulenko},   
\author[IHEP]{Y.M.~Melnick},   
\author[IHEP]{A.P.~Meschanin},   
\author[IHEP]{N.E.~Mikhalin},   
\author[IHEP]{N.G.~Minaev},   
\author[IHEP]{V.V.~Mochalov},   
\author[IHEP]{D.A.~Morozov},   
\author[IHEP]{L.V.~Nogach},   
\author[IHEP]{A.V.~Ryazantsev\thanksref{addr}},
\thanks[addr]{corresponding author, email: ryazantsev@mx.ihep.su}   
\author[IHEP]{P.A.~Semenov},   
\author[IHEP]{V.K.~Semenov},   
\author[IHEP]{K.E.~Shestermanov},   
\author[IHEP]{L.F.~Soloviev},   
\author[SYR]{S.~Stone},   
\author[IHEP]{A.V.~Uzunian},   
\author[IHEP]{A.N.~Vasiliev},   
\author[IHEP]{A.E.~Yakutin},   
\author[FNAL]{J.~Yarba}   
\collab{BTeV electromagnetic calorimeter group}   
\date{\today}   
   
\address[IHEP]{Institute for High Energy Physics, Protvino, Russia}   
\address[FNAL]{Fermilab, Batavia, IL 60510, U.S.A.}      
\address[UMN]{University of Minnesota, Minneapolis, MN 55455, U.S.A.}   
\address[SYR]{Syracuse University, Syracuse, NY 13244-1130, U.S.A.}   
   
\begin{abstract}   
Radiation damage in lead tungstate crystals reduces their transparency. 
The calibration that relates the amount of light detected in such crystals 
to incident energy of photons or electrons is of paramount importance to
maintaining the energy resolution the detection system. We report on 
tests of lead tungstate crystals, read out by photomultiplier tubes, 
exposed to irradiation by monoenergetic electron or pion beams. 
The beam electrons themselves were used to measure the scintillation 
light output, and a blue light emitting diode (LED) was used to 
track variations of crystals transparency.
We report on the correlation of the LED measurement with radiation damage 
by the beams and also show that it can accurately monitor the crystals 
recovery from such damage.
\end{abstract}

   
\end{frontmatter}   
   
\section{Introduction}   
Electromagnetic calorimeters built of  
the lead tungstate (PbWO$_4$, PWO) scintillating crystals 
will be used in several high energy physics experiments, such 
as ALICE and CMS at the CERN LHC~\cite{nimpwo,alice,cms}.
This work was done for studies of the electromagnetic calorimeter for 
BTeV at the FNAL Tevatron Collider~\cite{prop}. 
Unfortunately, BTeV was terminated by the U.S. Dept. of Energy. 
  
Over the last several years, the PWO crystals have been extensively studied  
at the Institute for High Energy Physics (IHEP) in Protvino,
Russia~\cite{nim0,nim1,nim2,nim3,nim4,nim5}.  
In particular, the studies confirmed that the PWO light output degrades 
under irradiation by high-energy electron or pion beam of high intensity.  
The light loss has a tendency to exhibit saturation when
the dose rate is kept at a constant level.
On the other hand, the light output changes whenever the radiation rate changes.    
Thus, they have to be monitored continuously during  
operation to maintain excellent energy and space resolutions of the calorimeter. 
 
Monte-Carlo study indicates that there would be enough electrons and positrons 
from photon conversion near the interaction region and semileptonic $B$ decays  
to calibrate the detector in-situ.
The rates of collecting electrons/positrons for the calibration vary in different 
areas of the calorimeter, ranging from less than 1 hour to about several hours of 
data taking to achieve the calibration accuracy of $(0.2-0.3)\%$.
Any changes in the crystal light output over this time scale must be monitored using 
the transparency measurements.
On the other hand, the transparency monitoring does not have to be stable over much 
longer time scale like a month since the electron calibration takes care of that part. 

We presume that the PWO light output degradation is caused by the changes 
in the transparency since no evidence of the scintillating mechanism damage has been 
found so far ~\cite{nim6}.  
Thus, a highly stable reference light pulser sending light through the crystal 
can be used to measure the transparency changes and predict the scintillation light 
output changes from the PWO crystals. 
 
When the coefficient of light absorption changes by $\Delta\alpha$,  
the light transmission changes by a factor of $e^{-\Delta\alpha\lambda}$,  
where $\lambda$ is the path length that light must travel from the source to the  
light detector.
It is expected that $\Delta\alpha\lambda<0.1$
for several hours of data taking, and it is relevant to use a linear
approximation. Therefore the fractional loss is given by
$\Delta\alpha\lambda$.
Considering that the path lengths for scintillation and  
monitoring light are different, the fractional losses for the two processes will  
be different, but are expected to be proportional.  

The dose rate profiles induced by electron and pion beams
are significantly different longitudinaly~\cite{nim3}. 
This could potentially lead to different proportionality constants for electron  
or pion irradiation.
Our studies addressed those issues for different crystals and under different  
conditions. A dedicated test beam run took place at the IHEP-Protvino  
test beam facility~\cite{nim1} in November-December 2002. 
  
Irradiation of PWO crystals has been done using high-intensity 
34~GeV pion beam and 23~GeV electron beam. Following the run, the crystals 
were left for a long-term natural recovery for over a 3-months period.  
Their light transmittance changes were measured with the LED monitoring 
system~\cite{nim4}. Results are presented in this paper.   
 
\section{Test beam apparatus and irradiation procedure} 
The IHEP-Protvino test beam facility is described in details in~\cite{nim2}. 
The major components are momentum tagging system and a prototype of the PWO  
crystal calorimeter. 
The prototype is a 5$\times$5 matrix of crystals from two vendors, Bogoroditsk and Shanghai, 
and is installed in a thermo-stabilized light-tight box on a moving platform. 
All the crystals are rectangular in shape, with a 27$\times$27 mm$^2$ cross-section  
and a 220 mm length. They were instrumented with 6-stage R5380Q Hamamatsu  
photomultiplier tubes (PMT).  
 
Before irradiation, scintillation light output of the crystals was measured  
with the use of a low intensity electron beam. All results presented in this paper
are normalized to the results of this very first calibration.

All crystals received from 500 rad to 1.5 krad of integral dose over the entire studies. 
During the electron irradiation runs, position of the electron peak itself was used 
to monitor the light output continuously.
During the irradiation by pions, data taking runs alternated with calibration runs 
by low intensity electron beam, to monitor changes in the crystals light output. 
For the pulse height analysis, only electrons that hit the central part  
(2$\times$2 mm$^2$) of the crystal's front face were selected using the data from  
drift chambers. 
 
Light transmittance change in the crystals was measured continuously with the 
use of a blue (470 nm) LED. Optical fibers guided light from the LED to the front side 
(opposite from the end where the PMT was attached) of the crystals. 
The typical path length of LED light in the crystal approximately equals the  
length of the crystal. The light comes out of the optical fiber with a  
characteristic full angle spread of 25$^\circ$; this angle is reduced to 
11$^\circ$ as the light enters the crystal from air. Thus, the path length of  
the light in a crystal should be increased by $1/\cos{11^\circ}$, i.e. order of 2\%.  
As for the scintillation light from incident particles, taking into  
account that this light is emitted isotropically and the crystal is wrapped
with Tyvek (diffuse reflecting material), its average path length to the 
photocathode is longer due to the multiple reflections. 
In addition, the LED system monitors the transparency of the crystal at 
a specific wavelength (in our case, 470~nm was chosen partially due 
to the availability of blue LEDs) and thus does not sample the 
entire spectrum of scintillation light.  The radiation damage 
effect is less severe at 470~nm than at 430~nm, which is the center  
of the PWO scintillation emission peak.
From these considerations, we expect that the ratio, $K$, of the light loss factors 
for the LED signal and the particle signal should be less than 1. 
 
\section{Experimental results}   
\subsection{Data from calibration runs} 
\begin{figure}   
\centering   
\includegraphics[width=0.75\textwidth]{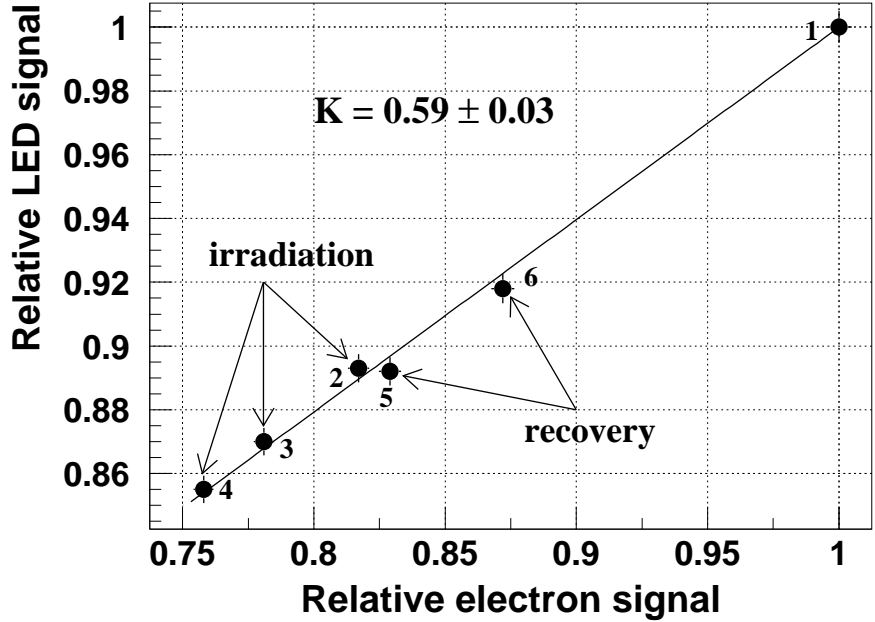}   
\caption{Blue LED {\em vs.} scintillation signal relative changes under  
pion irradiation for Shanghai T9 crystal. Points representing the data  
of the calibration runs are fitted to a straight line. The slope, $K=0.59$, 
was obtained with an accuracy of~$\pm5\%$.}   
\label{fig:pix2y4_LED-electron}   
\end{figure}

The mean pulse heights of the scintillation signal and of the LED signal  
were obtained using the data from the calibration runs and normalized to the 
results of the very first calibration. 
Fig.~\ref{fig:pix2y4_LED-electron} shows an example of the correlation between 
the relative changes of the LED signal {\em vs.} relative changes of the scintillation 
signal.
Points 1--4 represent measurements taken during the pion irradiation period; 
they fit very well to a linear function. 
Points 5--6 were taken when the crystal started to recover. 
By this we mean that the
high intensity pion beam moved away from this crystal and onto other crystals,
thus the dose rate on the spot decreased significantly and the light output 
started to restore, as was measured in the subsequent calibration runs.
It has to be noted that points 5--6 agree very well with the same linear fit 
applied to the data taken during the irradiation period.  
The fit function is shown below: 
\begin{equation} 
1-y = K(1-x)~,  
\label{eq:correlation} 
\end{equation} 
 
where $x$ and $y$ are relative electron and LED signals, respectively. 
For this particular crystal, the proportionality  
coefficient was obtained to be $K=0.59$ with an accuracy of~$\pm5\%$.   
    
\begin{figure}   
\centering   
\includegraphics[width=0.75\textwidth]{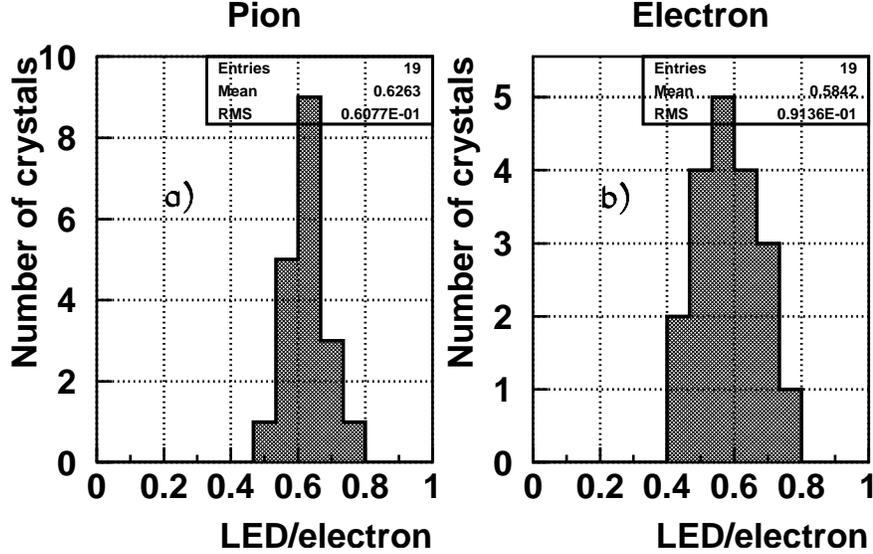}   
\caption{Linear fit coefficients between the blue LED and electron   
signals as a result of (a) pion irradiation; (b) electron irradiation.}   
\label{fig:coeff_pi19} 
\end{figure}

The same calculations were applied for all the crystals that have been 
irradiated either by electrons or pions.  
Distributions of the coefficients obtained from a linear fit of the 
LED {\em vs.} electron dependencies for each crystal are presented in  
Fig.~\ref{fig:coeff_pi19} for (a) pion and (b) electron irradiation. 
The mean values of the two distributions are the same within errors.
  
\subsection{Continuous electron calibration} 
 

All electrons incident within 3$\times$3 array of the crystals and satisfying the
condition, that the energy deposit over 9 crystals in this array was within $\pm 10\%$
of the beam energy, were selected for this analysis.   
The electron irradiation data were subdivided into smaller data sets, each set 
corresponded to 15 min of data taking. The mean signals for the 9 crystals in 
the array were calculated for each of the subsets. A standard inverse matrix iteration
procedure of crystal calibration required not more than 6 iterations. 

\begin{figure}   
\centering   
\includegraphics[width=0.85\textwidth]{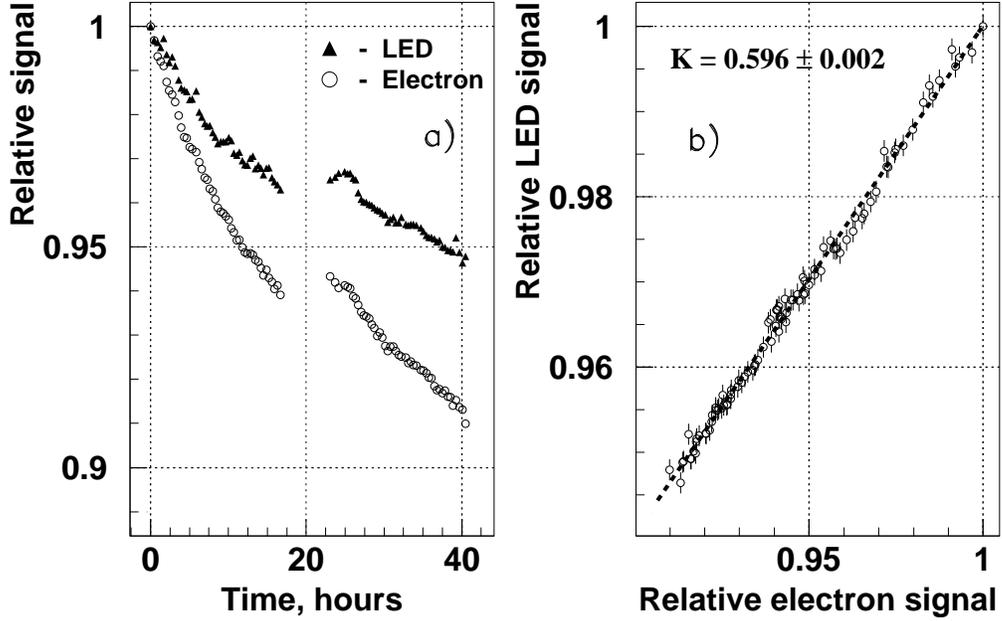}   
\caption{(a)The dependence of electron and blue LED normalized    
signals on time and (b)their correlation for Shanghai T9 crystal 
during the electron irradiation period.}   
\label{fig:elx2y4_all}   
\end{figure}

Fig.~\ref{fig:elx2y4_all}(a) shows the electron signal and the blue LED signal {\em vs.} 
time for one of the crystals during the electron irradiation period when the average 
dose rate was 20 rad/h; 
both signals have been normalized to the light output measured at the beginning of 
the irradiation period. 
Fig.~\ref{fig:elx2y4_all}(b) shows the correlation between relative electron signal 
and relative blue LED signal. 
The linear fit coefficient $K=0.596$ is computed with much better accuracy 
of $\pm 0.3\%$ than in the case of discrete calibration runs in the pion irradiation data.  

\begin{figure} 
\centering 
\includegraphics[width=0.95\textwidth]{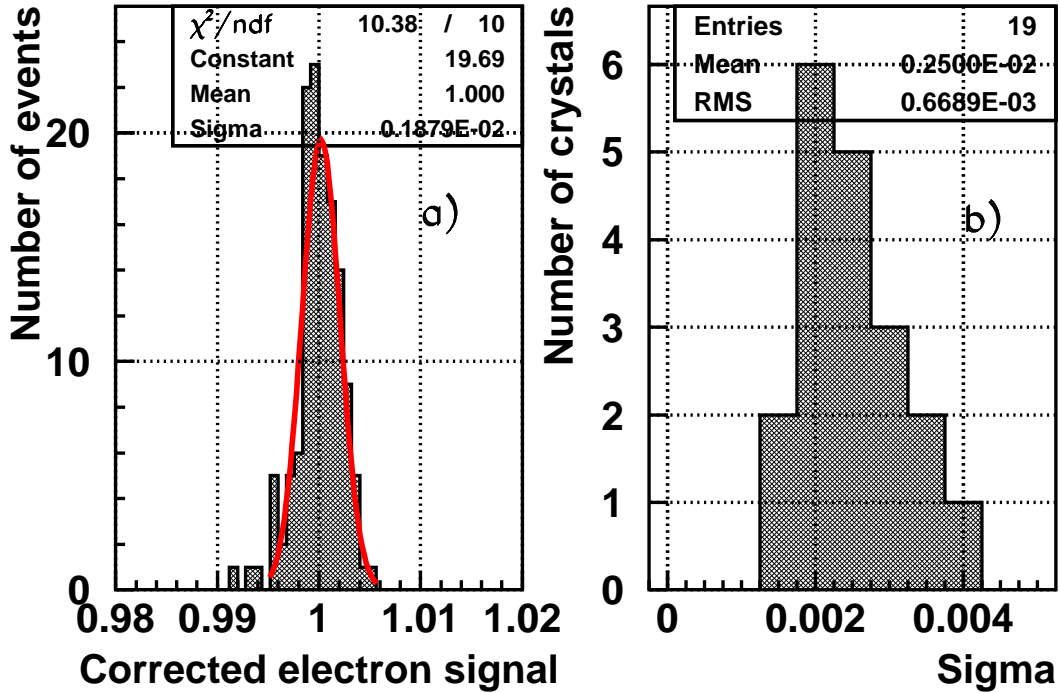} 
\caption{(a) Distribution of the electron signals corrected for the LED signals
over 35 hours of continuous crystal irradiation fitted by Gaussian ($\sigma$=0.2\%). 
(b) Distribution of Gaussian sigma values for 19 crystals.} 
\label{fig:el24_corr} 
\end{figure}

Fig.~\ref{fig:el24_corr} demonstrates how accurate the energy correction can be
in a single crystal with the use of a stable light calibration source if the 
coefficient $K$ in formula~\ref{eq:correlation} is known.  
The results presented here were obtained over 35 hours of data taking.
While the response changed over each 15-minutes period, the effect was corrected
according to the change in the LED signal and with the knowledge of the linear fit
coefficient that is shown in Fig.~\ref{fig:elx2y4_all}(a). 
The corrected energy distribution fitted by Gaussian has $\sigma$ equal to 0.2\%.
Fig.~\ref{fig:el24_corr}(b) shows the distribution of Gaussian $\sigma$ computed for 19
crystals with mean at 0.25\% and r.m.s. about 0.07\%.

\section{Long time crystals recovery} 
\begin{figure} 
\centering 
\includegraphics[width=0.75\textwidth]{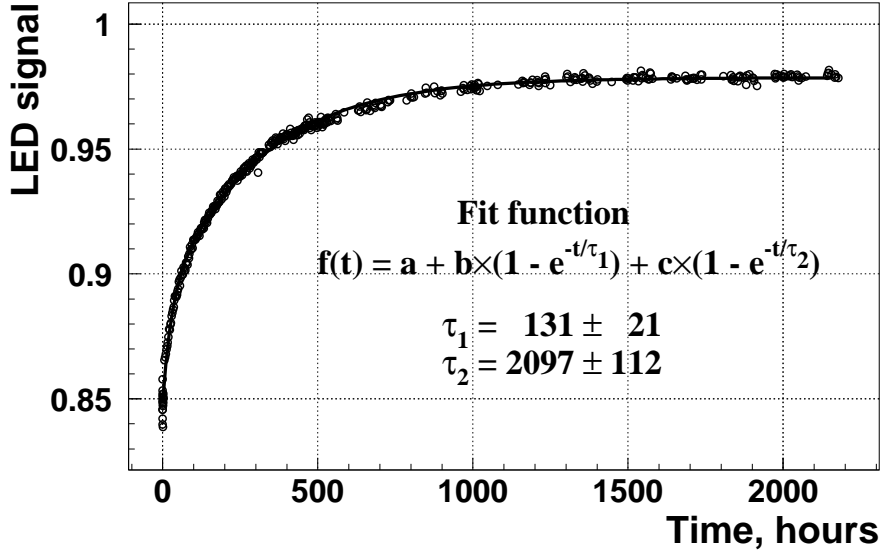} 
\caption{Dependence of the blue LED signal on time over 2200 hours of 
natural crystal's recovery. The parameters a,b and c are as follows : a=0.855$\pm$0.005,
b=0.097$\pm$0.06, c=0.048$\pm$0.016.} 
\label{fig:rec_time} 
\end{figure}

We observed that the transparency of the crystals recovered upwards to its 
level before irradiation.
Light output from the crystals was constantly monitored with the blue LED 
during for more than 3 months. 
Fig.~\ref{fig:rec_time} shows typical recovery process for one of the crystals. This
crystal was irradiated with the dose rates which varied from 15 to 30~rad/h 
and accumulated 2.2~krad absorbed dose. 
 
The experimental results were fitted with function: 
 
\begin{equation} 
f(t)=a+b\times(1-e^{-t/\tau_1})+c\times(1-e^{-t/\tau_2}). 
\label{eq:fitfunction} 
\end{equation} 
 
\begin{figure}   
\centering   
\includegraphics[width=0.75\textwidth]{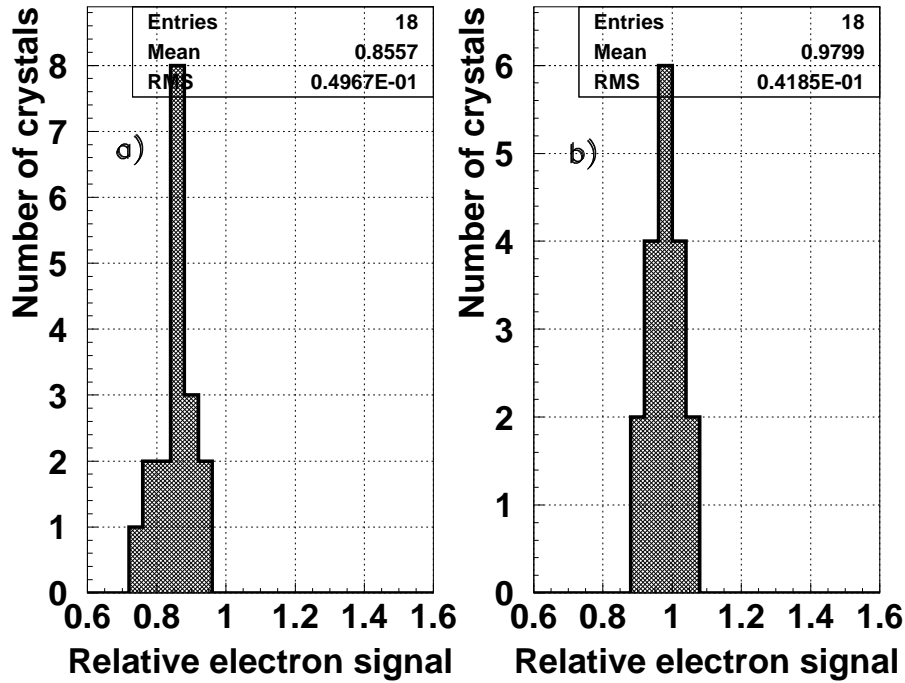}   
\caption{(a) after irradiation; (b) after the 2200 hours recovery.}
\label{fig:recovery}   
\end{figure}

Besides continuous monitoring of the changes in the crystals light transmittance
with the LED, 
at the end of the recovery period of more than 2200 hours, a calibration run 
with a low intensity electron beam was taken.   
The results of the calibration run were compared to those from the crystals  
calibration at the end of the irradiation run.  
Fig.~\ref{fig:recovery} shows that on average the light output from the 
crystals degraded to 86\% of its initial level at the end of the  
irradiation run but naturally recovered up to 98\%.
 
\section{Conclusions} 
The goal of this study was to confirm that the electromagnetic calorimeter made of
lead tungstate crystals read out by photomultiplier tubes  
can be continuously calibrated to the required accuracy with the use of LED-based 
monitoring system within a period of 1 day or shorter.  
 
We studied crystals behaviour under electron or hadron irradiation and whether the 
changes in their responses to electrons would scale well with their response to the blue LED. 
 
We found that the relative changes of the LED and electron signals can be approximated 
by a linear function in both the electron and pion irradiation studies. The obtained linear 
fit coefficients are consistent with each other.
This strongly suggests that in the real experimental environment, where the crystals 
will be irradiated by a mixture of hadrons, gammas and electrons, linear fit will be 
sufficient for the calorimeter's calibration. 
 
 

When the electron data were corrected for the transmission loss, which is due to irradiation, 
using the LED data, the corrected energy measurements are constant to within  $\pm 0.25\%$.
This satisfies one of the most important technical requirements of modern experiments.  
 
Over the 3-months long recovery period that followed the irradiation run we found that 
the light output of the crystals restored from an average of 86\% to 98\%.
It was also found that, for a given crystal, correlations between electron signal and 
blue LED  signal are linear and are the same if measured during irradiation or 
during the recovery period.  
   
\section{Acknowledgements}   
This work was supported by the U.S. National Science  
Foundation and the Department of Energy as well as the Russian  
Foundation for Basic Research grant 02-02-39008.

\end{document}